# Assessing Validity of ICD-10 Administrative Data in Coding Comorbidities

Jie Pan[1,2,*], PhD; Seungwon Lee[1,2,3], MPH PhD; Cheligeer Cheligeer[1,3], PhD; Bing Li[1,3], MA; Guosong Wu[1,2], PhD; Catherine A Eastwood[1,2], RN, PhD; Yuan Xu[1,2,4,5], MD, PhD; Hude Quan[1,2] MD, PhD

[1] Centre for Health Informatics, Cumming School of Medicine, University of Calgary, Calgary, AB, Canada

[2] Department of Community Health Sciences, Cumming School of Medicine, University of Calgary, Calgary, AB, Canada

[3] Provincial Research Data Services, Alberta Health Services, Edmonton, AB, Canada

[4] Department of Oncology, University of Calgary, Tom Baker Cancer Centre, Calgary, AB, Canada

[5] Department of Surgery, Foothills Medical Centre, University of Calgary, Calgary, AB, Canada

# Abstract

**Objectives**: Administrative data is commonly used to inform chronic disease prevalence and support health informatics research. This study assessed the validity of coding comorbidity in the International Classification of Diseases, 10th Revision (ICD-10) administrative data.

**Methods**: We analyzed three chart review cohorts (4,008 patients in 2003, 3,045 in 2015, and 9,024 in 2022) in Alberta, Canada. Nurse reviewers assessed the presence of 17 clinical conditions using a consistent protocol. The reviews were linked with administrative data using unique identifiers. We compared the accuracy in coding comorbidity by ICD-10, using chart review data as the reference standard.

**Results**: Our findings showed that the mean difference in prevalence between chart reviews and ICD-10 for these 17 conditions was 2.1% in 2003, 7.6% in 2015, and 6.3% in 2022. Some conditions were relatively stable, such as diabetes (1.9%, 2.1%, and 1.1%) and metastatic cancer (0.3%, 1.1%, and 0.4%). For these 17 conditions, the sensitivity ranged from 39.6-85.1% in 2003, 1.3-85.2% in 2015, and 3.0-89.7% in 2022. The C-statistics for predicting in-hospital mortality using comorbidities by ICD-10 were 0.84 in 2003, 0.81 in 2015, and 0.78 in 2022.

**Discussion**: The under-coding could be primarily due to the increase of hospital patient volumes and the limited time allocated to coders. There is a potential to develop artificial intelligence methods based on electronic health records to support coding practices and improve coding quality.

**Conclusion**: Comorbidities were increasingly under-coded over 20 years. The validity of ICD-10 decreased but remained relatively stable for certain conditions mandated for coding. The under-coding exerted minimal impact on in-hospital mortality prediction.

## What is known on this topic?

- The quality of coding comorbidity conditions in administrative health data contributes to downstream health research and population surveillance.
- Many health data validation efforts have been initiated and implemented by Canadian healthcare systems.

## What this study adds?

- The comorbidities are increasingly under-coded over 20 years, but certain conditions remain stable.
- The prediction of in-hospital mortality is minimally affected by the under-coding of comorbidities.

## How this study might affect research, practice or policy?

- The decreasing accuracy highlights the need for improved methodologies, such as analyzing clinical notes with advanced machine learning models, in health data research.
- The under-coding of comorbidities calls for policy adjustments, such as allocating more manpower or developing automation tools, to enhance coding practices for more accurate healthcare reporting.

## Introduction

Canada has an extensive collection of electronic health databases that contribute to system planning and health research. These databases include administrative health databases, clinical registries, electronic health records, and health surveys.[1,2] Administrative health databases collect information for healthcare operations, system performance, and population surveillance, among other activities. One example database, the discharge abstract database (DAD), includes administrative, clinical, and demographic information from patient encounters at acute care facilities. The provinces collect and code the data using the International Classification of Diseases, Tenth Revision (ICD-10), Canadian edition (ICD-10-CA), and then submit the details to the Canadian Institute for Health Information (CIHI). Such administrative health databases enable analytics on disease prevalence and healthcare utilization patterns. The collected administrative health data is crucial in supporting significant public health initiatives like the Canadian Chronic Disease Surveillance System (CCDSS)[3].

Comorbidity indices like the Charlson Comorbidity Index[4] and the Elixhauser Comorbidity Index[5] are routinely used to quantify the burden of comorbid conditions on administrative health data. Quan *et al.*[6] assessed the validity of coding comorbidities using the ICD-10 codes in 2008, found under-reported issues for 31 conditions, and reported a similar performance on using ICD-9 codes. Since then, many efforts have been made to ensure the reliable and accurate information used for research and healthcare system monitoring, including the data quality frameworks on the administrative data[7,8], computational assessment of data quality[9], and the analysis of quality barriers in health systems[10]. Despite these validation efforts, little is known about whether the coding quality of comorbidity is consistent or has changed over time.

Since administrative health databases (e.g., DAD) inform chronic disease prevalence and population health status, we designed this study to assess the validity of ICD-10-CA for coding comorbidities in acute care settings and to determine whether there were changes in validity over the years. To this aim, we assessed the prevalence of comorbidities and in-hospital mortality prediction between a series of chart review data and the originally coded ICD-10-CA administrative data from 2003 to 2022. This permitted us to examine trends in coding accuracy, changes in coding guidelines, and the potential need for data quality initiatives.

## Methods

We conducted a retrospective cohort study utilizing three previously collected chart review databases from 2003 to 2022. The datasets were undertaken at the four acute care facilities in Alberta, Canada.

### Chart Review and Discharge Abstract Databases

The Discharge Abstract Database is a national administrative database that includes demographic and clinical details from inpatient hospitalization and encounters. CIHI maintains the DAD with data submitted from provincial authorities. CIHI sets the national standard for training, and provincial authorities (e.g., Alberta Health Services in Alberta, Canada) are responsible for data collection. Currently, professionally trained health information management clinical coding specialists review the raw medical charts post-discharge of a patient encounter and assign ICD-10-CA diagnosis codes for the record.[11] DAD can include up to 25 ICD-10-CA diagnosis codes, including both primary and secondary codes. All the diagnosis codes were included for validity analysis in this study.

The three chart review cohorts were assembled in 2003[6], 2015[12], and 2022[13], with sample size calculated at α = 0.05 and power = 0.8 for all cohorts. Specifically, three adult patient cohorts were discharged between January 1, 2003, and June 30, 2003 (2003 cohort), January 1, 2015, and June 30, 2015 (2015 cohort), and January 1, 2017, and June 30, 2022 (2022 cohort).

The chart review process was conducted by trained nurse reviewers, with two reviewers in the 2003 cohort and six in the 2015 and 2022 cohorts. Registered nurses specializing in surgery, general internal medicine, intensive care, oncology, or cardiology and possessing at least three years of clinical experience were recruited for chart review. They received extensive training to independently review inpatient electronic hospital records, including cover pages, discharge summaries, trauma and resuscitation records, admission, consultation, diagnostic, surgery, pathology, and anesthesia reports, and multidisciplinary daily progress notes. Reviewers followed a standardized protocol based on Charlson comorbidity definitions[14], undergoing training, agreement studies, and full chart examinations. Inter-rater reliability, assessed using Kappa statistics, demonstrated strong agreement across conditions, with the 2003 cohort and 2015 cohort achieving Kappa > 0.80 for 17 conditions and the 2022 cohort showing Kappa > 0.72 for 15 conditions. This consistent review process ensured high data quality across all cohorts.

The summary of detailed data extraction and chart review procedure for three cohorts can be found in Supplementary Table S1. The detailed review process for the 2003 cohort can be found in Quan *et al.*'s work[6]. Eastwood *et al.*[12] and Wu *et al.*[13,15] described the review process for the 2015 cohort and 2022 cohort and their reference standards of comorbidities, respectively. The reviewed charts were then linked with patients' administrative data using a unique lifetime identifier, chart number, and admission date.

The chart review process was consistent across the three selected cohorts. However, between the 2015 and 2022 cohorts, slight variations in the chart review reference standards existed for two conditions: peptic ulcer disease and depression. These subtle changes in definitions were based on the evolving evidence of clinical practices and their guidelines.

Statistical analysis

The prevalence of 17 comorbidities was defined by Quan's earlier work that established the ICD-10-CA code algorithms[14]. The algorithms' results were compared against the chart review labels set as reference standards for the respective databases. Sensitivity, specificity, positive predictive value (PPV), and negative predictive value (NPV) were calculated. We built logistic regression models using 17 comorbidities as independent variables to examine the validity over time in predicting in-hospital mortality. The C-statistics were calculated for each model to investigate the prediction performance.

Existing studies demonstrated that including prior years' administrative data could ascertain chronic cases. [16,17] To assess whether volume change of data could improve the performance, we extracted additional 1, 3- and 5-years prior DAD data on the 2022 cohort. Each patient's prior data was extracted from the date of admission record and linked. We then recalculated the prevalence and the performance of the comorbidities.

*Figure 1 Difference between chart reviewer and ICD-10-CA coding for comorbidities across three cohorts (2003, 2015, 2022). The differences are calculated as chart data prevalence minus ICD-10-CA coding prevalence. Each line represents a different comorbidity, showing how the discrepancy has evolved over time. Six conditions (peptic ulcer disease, hypertension, cancer, depression, chronic pulmonary disease, and renal disease) were highlighted with solid lines because of their significant changes; the remaining conditions were represented by dashed lines.*

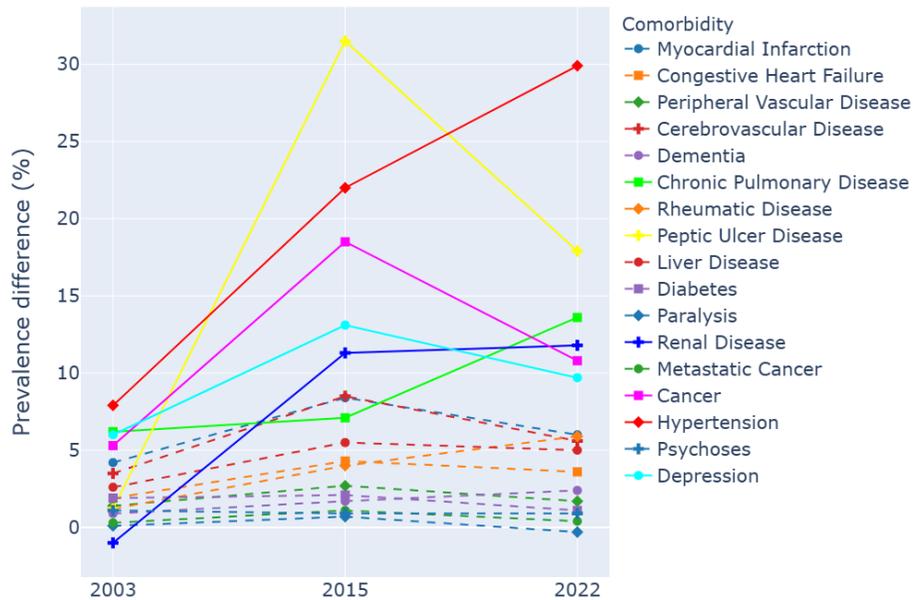

## Results

### Prevalence of comorbidities

The prevalence of 17 Charlson comorbidities over the years (2003, 2015, and 2022) are shown in Table 1. The difference in prevalence identified by chart reviewers and ICD-10-CA in DAD was calculated within each cohort, as shown in Figure 1. We highlighted six conditions (peptic ulcer disease, hypertension, cancer, depression, chronic pulmonary disease, and renal disease) with the most significant changes in prevalence between ICD-10 coding and chart review, using solid lines. These same conditions were highlighted in other figures, allowing for easier tracking of changes in performance. The mean difference in prevalence between chart reviews and administrative data for these 17 conditions was 2.1% in 2003, 7.6% in 2015, and 6.3% in 2022. Fifteen conditions were consistently under-reported by ICD-10 over the years and 2 conditions were slightly over-reported (renal disease in 2003 and paralysis in 2022). The coding of 5 comorbidities (dementia, chronic pulmonary disease, rheumatic disease, renal disease, and hypertension) was increasingly under-reported. Three comorbidities had relatively consistent coding prevalence and difference, such as peripheral vascular disease (1.4%, 2.7%, and 1.7%), diabetes (1.9%, 2.1%, and 1.1%), and metastatic cancer (0.3%, 1.1%, and 0.4%). Two conditions (peptic ulcer disease and cancer) had consistent ICD-coding prevalence but large differences in chart data.

Table 1 Comparison of comorbidity prevalence by chart reviewer and ICD-10-CA coding across three cohorts (2003, 2015, 2022).

| Comorbidity | 2003 cohort N = 4,008 | | 2015 cohort N=3,045 | | 2022 cohort N = 9,024 | |
|---|---|---|---|---|---|---|
| | Chart data (%) | ICD-10-CA (%) | Chart data (%) | ICD-10-CA (%) | Chart data (%) | ICD-10-CA (%) |
| Myocardial Infarction | 12.7 | 8.5 | 11.7 | 3.3 | 11.0 | 5.0 |
| Congestive Heart Failure | 8.3 | 6.4 | 11.3 | 7.0 | 9.7 | 6.1 |
| Peripheral Vascular Disease | 4.3 | 2.9 | 4.9 | 2.2 | 3.2 | 1.5 |
| Cerebrovascular Disease | 8.1 | 4.6 | 11.8 | 3.3 | 9.9 | 4.3 |
| Dementia | 3.3 | 2.4 | 5.7 | 4.0 | 4.7 | 2.3 |
| Chronic Pulmonary Disease | 15.0 | 8.8 | 14.7 | 7.6 | 17.8 | 4.2 |
| Rheumatic Disease | 2.6 | 1.4 | 5.0 | 1.0 | 6.2 | 0.3 |
| Peptic Ulcer Disease | 2.5 | 1.3 | 32.4 | 0.9 | 19.1 | 1.2 |
| Liver Disease | 5.0 | 2.4 | 7.8 | 2.3 | 6.9 | 1.9 |
| Diabetes | 14.6 | 12.7 | 19.2 | 17.1 | 18.9 | 17.8 |

| Paralysis | 1.5 | 1.4 | 1.7 | 1.0 | 0.4 | 0.7 |
| --- | --- | --- | --- | --- | --- | --- |
| Renal Disease | 4.0 | 5.0 | 14.2 | 2.9 | 13.9 | 2.1 |
| Metastatic Cancer | 4.4 | 4.1 | 4.6 | 3.5 | 5.6 | 5.2 |
| Cancer | 14.1 | 8.8 | 29.4 | 10.9 | 21.0 | 10.2 |
| Hypertension | 30.2 | 22.3 | 48.5 | 26.5 | 44.4 | 14.5 |
| Psychoses | 2.9 | 1.8 | 3.0 | 2.1 | 1.0 | 0.1 |
| Depression | 11.9 | 5.9 | 18.0 | 4.9 | 10.6 | 0.9 |

## Comorbidity coding performance

Using chart data as a reference standard, the coding performance of comorbidities in administrative data over the years 2003, 2015, and 2022, are shown in Figure 2. The sensitivity (ranged 39.6-85.1% in 2003, 1.3-85.2% in 2015, and 3.0-89.7% in 2022, see detailed data in Supplementary Table S2), showed a noticeable trend of decreasing for 16 conditions over time except for diabetes (85.1%, 85.2%, and 89.7%). The specificity remained high (ranged 97.8-99.8%, 98.1-99.9%, and 97.5-100.0%, respectively) across all conditions, which suggests a consistent ability to correctly identify non-cases.

*Figure 2 Trends in coding accuracy metrics for comorbidities by ICD-10-CA across three cohorts (2003, 2015, 2022). It calculated the sensitivity, specificity, positive predictive value (PPV), and negative predictive value (NPV) for various comorbidities using chart data as reference standards. Six conditions (peptic ulcer disease, hypertension, cancer, depression, chronic pulmonary disease, and renal disease) were highlighted with solid lines because of their significant changes; the remaining conditions were represented by dashed lines.*

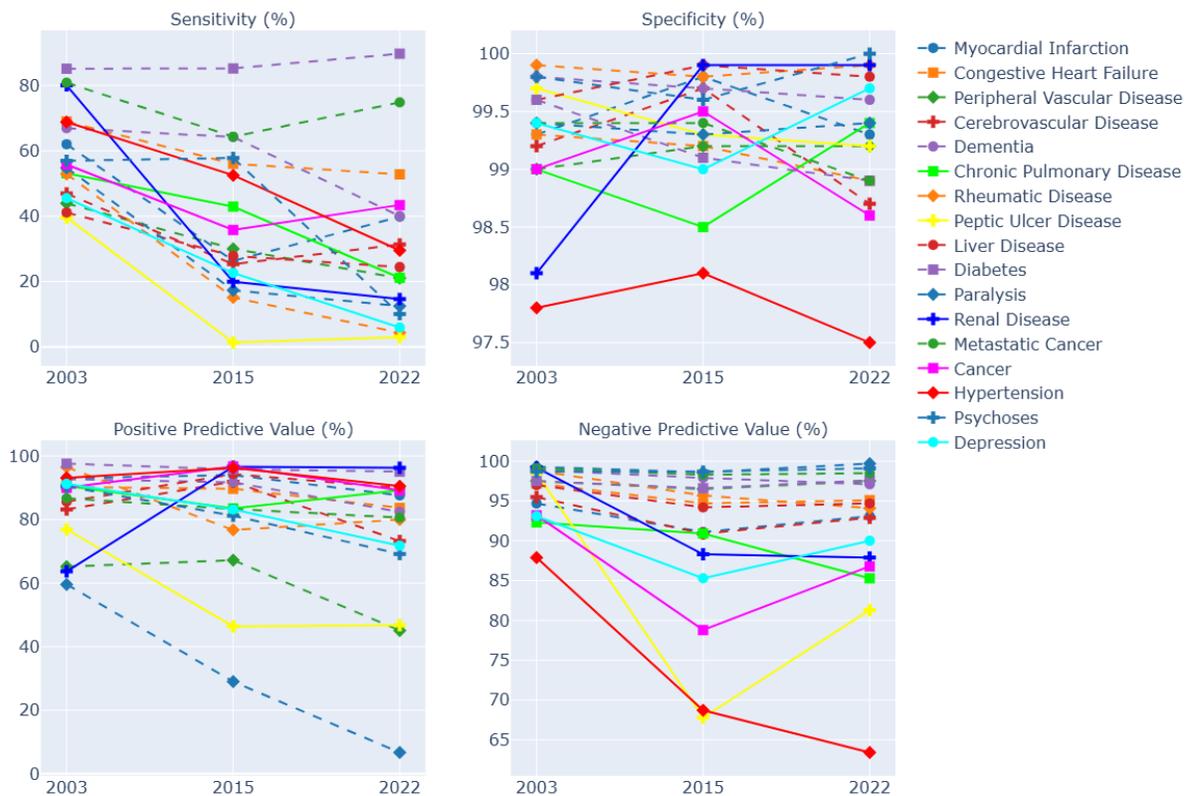

For NPVs, most conditions were generally high across years, suggesting substantial accuracy in identifying non-cases, but two conditions, chronic pulmonary disease (92.3%, 90.9%, and 85.3%) and hypertension (87.9%, 68.7%, and 63.4%), had declined over time. PPVs showed that most conditions had a slight decrease in accuracy, while two conditions, peptic ulcer disease (76.9%, 46.4%, and 46.8%) and paralysis (59.6%, 29.0%, and 6.7%), had a significant drop. The detection of renal disease had increased from 63.7% to 96.6% and 96.3%. Some conditions had relatively consistent PPVs, such as diabetes (97.6%, 95.8%, and 95.1%) and metastatic cancer (86.7%, 83.3%, and 80.6%).

## Coding performance with the addition of prior years of administrative data

The prevalence of the comorbidities after linking with prior years' data is shown in Figure 3(a). The prevalence of 17 conditions increased with more data included when compared to the original prevalence. Details are available in Supplementary Table S3. Specifically, hypertension (14.5% in 2022, 21.2% with one prior year data, 25.0% with three prior years data, and 26.6% with five prior years data) had the most increase in prevalence from its original dataset. Four conditions had a higher prevalence than chart data after including five prior years of administrative data, such as congestive heart failure, diabetes, paralysis, and metastatic cancer. We also calculated the sensitivity, specificity, PPV, and NPV on the linked datasets, as shown in Supplementary Table S4. The results were plotted together with the coding performance in 2022 cohorts, as shown in Figure 3(b). The sensitivity and NPV improved by including more data across 17 Charlson conditions. For example, the NPV of hypertension experienced a significant increase (63,4% in 2022, 66.4% with one prior year data, 69.5% with three prior years data, and 71.2% with five prior years data). The specificity and PPV of most conditions were slightly declined. Two conditions (cancer and hypertension) had noticeable drops in specificity. Paralysis (6.7% in 2022, 7.3%, 11.3%, and 12.2%) had an improvement in PPV.

*Figure 3 Prevalence (a) and coding performance (b) changes of comorbidities by ICD-10-CA from 2022 cohort to 2022 cohort with additional 1 prior year, 3 prior years, and 5 prior years of administrative data. (b) It calculated the sensitivity, specificity, positive predictive value (PPV), and negative predictive value (NPV) for various comorbidities using chart data of the 2022 cohort as reference standards. Six conditions (peptic ulcer disease, hypertension, cancer, depression, chronic pulmonary disease, and renal disease) were highlighted with solid lines because of their significant changes; the remaining conditions were represented by dashed lines.*

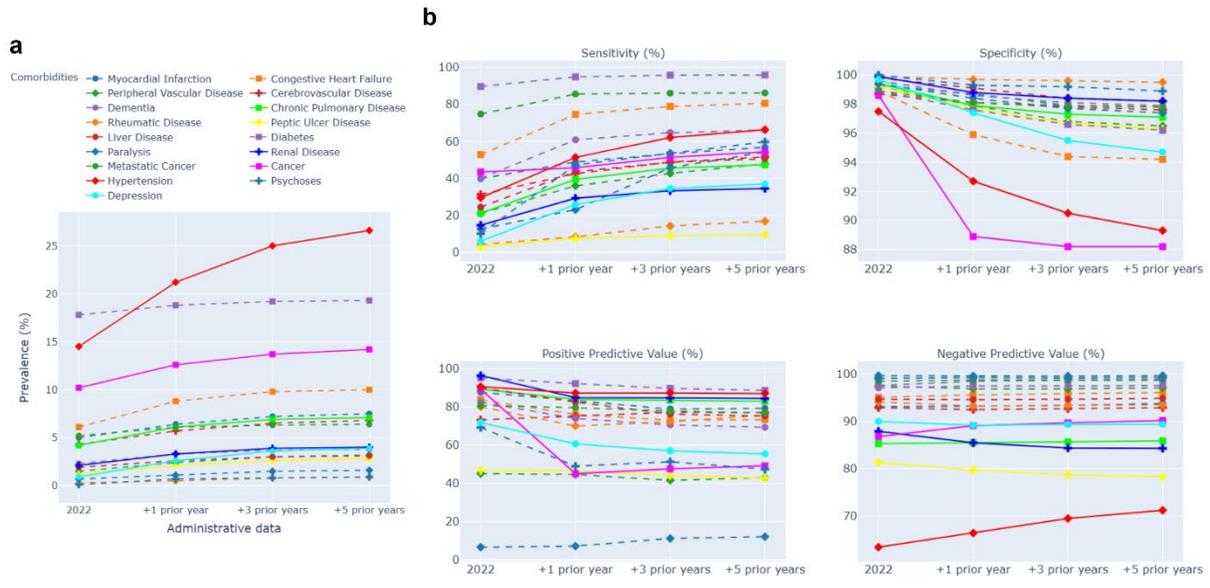

## Predicting in-hospital mortality based on comorbidities

We examined the efficacy of predicting in-hospital mortality using the identified comorbidities in 2003, 2015, and 2022 and the augmented administrative data for the 2022 cohort, as shown in Table 2. The mortality per dataset was also calculated. The mortality rates were similar in the 2003 and 2015 cohorts (2.6% and 2.1%), whereas 2022 had a higher mortality rate (7.2%). The C-statistics for in-hospital mortality based on ICD-10 were 0.84 in 2003, 0.81 in 2015, and 0.78 in 2022, which were very close to those obtained from chart data. The C-statistics were slightly decreased for both data, but their predictive accuracy remained reasonably high (around 80%). Including an additional one prior year, three preceding years, and five preceding years of administrative data, the predictive accuracy of in-hospital mortality by ICD-10 codes was slightly improved from 0.783 to 0.79, 0.792, and 0.793.

*Table 2 C-statistics comparison for predicting in-hospital mortality using comorbities across three Cohorts (2003, 2015, 2022) and 2022 cohort with the addition of prior years of administrative data. A logistic regression model was built for each cohort using in-hospital mortality as the outcome and 17 comorbidities as predictors.*

|  | Chart data | ICD-10 | ICD-10 in 2022 |
| --- | --- | --- | --- |

|  | 2003 | 2015 | 2022 | 2003 | 2015 | 2022 | +1 prior year | +3 prior years | +5 prior years |
|---|---|---|---|---|---|---|---|---|---|
| C-statistics | 0.85 | 0.824 | 0.808 | 0.843 | 0.813 | 0.783 | 0.79 | 0.792 | 0.793 |
| Mortality rate (%) | 2.6 | 2.1 | 7.2 | 2.6 | 2.1 | 7.2 | 7.2 | 7.2 | 7.2 |

# Discussions

This study evaluated the in-hospital coding of Charlson comorbidities in DAD from four hospital sites against chart-reviewed databases spanning 20 years. We found that DAD increasingly under-coded conditions but remained relatively stable for certain conditions (e.g., diabetes, metastatic cancer, and peripheral vascular disease). The longitudinal variations in the coding quality of comorbidities exerted minimal impact on in-hospital mortality predictions.

The quality of administrative data is crucial to inform the health services policy, program planning, and public health stakeholders in downstream processes. In the Canadian context, CIHI and the Public Health Agency of Canada (PHAC), among other agencies, emphasize the importance of maintaining accuracy, timeliness, and usability in data quality assessments in their frameworks for continuously evaluating administrative health databases against these quality dimensions.[18,19] Aynslie Hinds et al.[2] reviewed the validation studies of administrative data and found that the common investigation subjects included case definitions for chronic conditions and diagnostic codes of comorbidity indices. While many studies, such as those by Smith et al. (2017)[7] and Iwig et al. (2023)[8], propose new tools for assessing data quality, there is limited evidence of their widespread implementation or evaluation of long-term results. A few studies have examined quality consistency over time.[7] Our study extended the literature by analyzing the consistency of coding practices over time within the same healthcare system, offering a focused examination of changes in coding accuracy for multiple comorbidities.

Compared to the work of Wei et al. (2020)[20], which reviewed existing validation studies on 81 conditions completed before 2014, our findings revealed a similar trend in conditions like diabetes coding accuracy, which remained relatively stable across years. For some other conditions, they reported hypertension sensitivity at 65%, while our 2015 cohort had a lower sensitivity of 53%; for congestive heart failure, the sensitivity ranged from 20% to 94%, while our results showed a decline from 69% in 2003 to 53% in 2022. These findings roughly align with our validation results. However, these results were obtained from different studies and diverse healthcare systems. By conducting our analysis within the same healthcare system over time, we could offer insights into the temporal changes in administrative data coding practices, providing a clearer picture of how comorbidity coding accuracy evolves and highlighting areas in need of improvement.

The prevalence difference analysis highlights the increasing under-reporting of several comorbidities, such as dementia, chronic pulmonary disease, and hypertension, over the years, and further highlights the present reality where the hospital system has experienced increased in-patient volumes and coders are required to meet this turnover. The work by Tang *et al.* [11] further demonstrated that there is a gap between clinician documentation and administrative coding process. This trend points to the need for improved communication between the stakeholders on the coding practices or the adoption of tools to support the coding process. Additionally, the consistent prevalence of certain conditions like diabetes and metastatic cancer in both chart data and ICD-10 coding underlines the reliability of these conditions in administrative data for epidemiological tracking.

The trends in coding performance over time provide valuable insights into the accuracy of capturing comorbidities. Overall, specificity remained high across all years for most conditions, indicating a consistent ability to correctly identify non-cases. However, sensitivity showed a noticeable decline for several conditions, such as myocardial infarction, congestive heart failure, and peripheral vascular disease, which suggests that the ability to detect true cases has weakened. For their detection, to include automation tools for analyzing existing data sources, such as electronic health records, could be beneficial. Regarding PPV, a decline was observed in certain conditions like peptic ulcer disease and paralysis, indicating a loss of accuracy in identifying true positives. Conversely, renal disease showed a substantial increase in PPV, from 63.7% in 2003 to 96.3% in 2022, which may reflect improved recognition for this condition. NPV remained relatively stable across most conditions. These findings highlight the need for ongoing evaluation of coding practices to ensure accurate identification of comorbidities in administrative data.

The large prevalence differences in conditions like chronic pulmonary disease and hypertension, with increasing under-reporting over time, directly affect performance metrics such as sensitivity, PPV, and NPV. The decreased prevalence in ICD-10 coding for these conditions leads to reduced sensitivity and PPV, as many true cases are missed. Consistent under-reporting can also affect NPV, overestimating the non-case population by misclassifying some true cases. In contrast, conditions with low prevalence differences, like peripheral vascular disease, diabetes, and metastatic cancer, have more stable metrics, as consistent coding accuracy better reflects the true disease burden.

The observed decline in the C-statistics for predicting in-hospital mortality over the years when using ICD-10 coding, relative to chart data, can be attributed to the differences in the prevalence and coding accuracy of comorbidities. Conditions such as chronic pulmonary disease and hypertension, which showed increasing under-reporting in ICD-10, led to a greater misclassification of true positive cases. They are critical comorbidities that significantly affect in-hospital mortality risk, resulting in a drop in C-statistics (from 0.84 in 2003 to 0.78 in 2022). On the other hand, conditions with more consistent coding practices, such as diabetes, contributed to relatively stable PPVs and a smaller impact on the overall predictive model.

Several factors contribute to the discrepancies in coding performance. Out of 17 conditions, diabetes and conditions that contributed most to prolonged stays in care facilities were mandated to be coded in the Canadian administrative database. Diabetes must be coded whenever it is documented, as outlined in the CIHI coding guidelines[21]. This requirement stems from the serious nature of diabetes and its potential for long-term complications affecting multiple systems in the body. Other conditions are only coded when they meet specific clinical criteria. This explains the relatively stable coding quality of diabetes and some other conditions. A few qualitative studies have indicated high barriers to achieving high-quality coding in the Canadian context[10,11], including incomplete documentation from providers, the requirement for faster turnaround time resulting in high pressure on coding specialists, and discrepancies in utilized terminologies between coding specialists and providers. The advent of the digital age has led to increased adoption of electronic health records in acute care facilities in recent years, compared to the early days in 2003. This increased adoption has likely resulted in a higher volume of data associated with coding. For example, the province of Alberta is near completion of implementing a province-wide clinical information system (i.e., Connect Care) incorporating EHR in all acute-care facilities throughout the province. Therefore, the above qualitative factors and changes in health system capacity could explain the decrease seen in this study.

The impact of including prior years' data was notable. The prevalence of comorbidities increased with the inclusion of earlier years' data, with hypertension showing the most significant increase. For this reason, many cohort selection algorithms typically use claims codes or hospitalization codes over two years to define chronic conditions.[22] Sensitivity and NPV were improved by including more data, whereas specificity and PPV slightly declined. The C-statistics for in-hospital mortality predictions decreased somewhat over the years but remained reasonably high (around 80%). This indicates the reliability of administrative data in providing accurate information for predicting patient outcomes, showcasing the robustness of healthcare analytics. After including data from prior years, the C-statistic remained stable, indicating that conditions in earlier episodes of care were not severe enough to impact in-hospital mortality during the latest hospitalization.

Our study has several limitations. First, our cohorts are restricted to four acute care facilities in Alberta. Hence, the results are not externally validated and require further verification. Second, the three datasets were collected at different time points and may be influenced by differing health systems and clinical practices. Last, this study is limited to inpatient data only, and as such, it may not be fully representative of outpatient coding practices. However, the study contains several strengths. First, the coding quality results are based on three separate chart review databases conducted on the same facilities spanning a long period of time. Second, the results are based on real-world evidence of system practices and can be helpful to inform the stakeholders.

Alberta Health Services Data & Analytics is currently implementing an infrastructure that can potentially implement large language models (LLM) into its arsenal of tools. Perhaps there exists a future where LLM can assist the coders with information extraction on comorbidities from the large volume of EHR data. Nevertheless, additional non-technical factors (e.g., documentation quality) require multi-disciplinary conversations and collaboration between all stakeholders (e.g., health system, physicians, coding specialists) to improve the coding quality of comorbidities. For instance, the quality of the DAD is closely tied to the quality of EHR data, as ICD coders can only use what physicians document in patient charts. Any gaps or inaccuracies in EHR data directly affect the quality of ICD coding. Many physicians may not realize the link between their documentation and ICD-based systems, highlighting the need for collaboration to enhance data accuracy. All of these will be explored in the future.

## Conclusion

The DAD increasingly included under-coded conditions but remained relatively stable for certain mandated conditions. The impact on in-hospital mortality prediction was minimal. Achieving high-quality coded data will require careful dialogues between all stakeholders while incorporating changes in health system infrastructure and clinical and health system practices.

## Funding Sources

This work was supported by Canadian Institutes of Health Research Operating Project Grants (201809FDN-409926-FDN-CBBA-114817).

## Competing Interests

Authors have no competing interests to disclose.

## Ethical Approval Statement

This study was approved by the Conjoint Health Research Ethics Board at the University of Calgary (REB19-0088 and REB-23-0535).

# Contributorship Statement

Jie Pan drafted the first manuscript and conducted the study design and analysis. Seungwon Lee assisted with data collection, manuscript writing, and results interpretation. Cheligeer Cheligeer assisted with data collection and modelling. Bing Li helped with data collection. Guosong Wu, Catherine A. Eastwood, and Yuan Xu provided interpretations of the study results. Hude Quan was the guarantor and responsible for the study design, and provided the interpretation framework of experimental results. All authors reviewed the manuscript from the perspectives of soundness, completeness, and novelty.